\documentclass[a4paper,12pt]{article}
\usepackage{indentfirst}
\usepackage{amsfonts,amsmath,amssymb,bm,a4wide,graphicx,cite,makeidx,multicol}
\usepackage[cp1251]{inputenc}

\usepackage {bm}

\begin{document}
\title{Spin light of electron in dense matter}

\author{Alexander Grigoriev~$^{a,b}$ \footnote{ax.grigoriev@mail.ru}, \
        Sergey Shinkevich~$^a$ \footnote{shinkevich@gmail.com}, \
        Alexander Studenikin~$^{a,b}$ \footnote{studenik@srd.sinp.msu.ru},
        \\
        Alexei Ternov~$^c$ \footnote {a\_ternov@mail.ru}, \ Ilya Trofimov~$^a$ 
        \\
   \small {\it $^a$Department of Theoretical Physics,}
   \\
   \small {\it Moscow State University, 119992 Moscow,  Russia }
   \\
   \small {\it $^b$Skobeltsyn Institute of Nuclear Physics,
   Moscow State University, 119992 Moscow, Russia}
   \\
   \small{\it $^c$Department of Theoretical Physics,}
   \\
   \small {\it Moscow Institute for Physics and Technology,
   141700 Dolgoprudny, Russia }}

\date{}

 \maketitle

\begin{abstract}

We derive the modified Dirac equation for an electron undergos an influence of
the standard model interaction with the nuclear matter. The exact solutions for
this equation and the electron energy spectrum in matter are obtained. This
establishes a rather powerful method for investigation of different processes
that can appear when electrons propagate in background matter. On this basis we
study in detail the spin light of electron in nuclear matter, a new type of
electromagnetic radiation which can be emitted by an electron moving in dense
matter.
\end{abstract}





\section{Introduction}
\par

The problem of particles interactions under an external environment influence,
provided by the presence of external electromagnetic fields or media, is one of
the important issues of particle physics. In addition to possibility for better
visualization of fundamental properties of particles and their interactions
being imposed by influence of an external conditions, the interest to this
problem is also stimulated by important applications to description of
different processes in astrophysics and cosmology, where strong electromagnetic
fields and dense matter may play an important role.

What is concerned the influence of strong external electromagnetic fields,
there is a well known method which enables one to account for an external field
influence on a charged particle exactly, rather than within the
perturbation-series expansion. In this techniques, known in quantum
electrodynamics as the Furry representation \cite{FurPR51}, consideration of a
quantum process is based on the use of exact solutions of the corresponding
modified Dirac equation for the particle wave function,
\begin{equation}\label{D_eq_QED}
\left\{ \gamma^{\mu}\big(i\partial_{\mu} -eA_{\mu}^{cl}(x)\big) - m_e
\right\}\Psi(x)=0,
\end{equation}
which accounts for the  external field classical potential $A_{\mu}^{cl}(x)$.
The quantized part of the potential $A_{\mu}^{q}(x)$, that corresponds to the
electromagnetic radiation  field, is treated within the perturbation-series
techniques.  A detailed discussion of this method can be found in
\cite{SokTerSynRad68}.

In a series of our papers \cite{StuTerPLB05,
GriStuTerPLB05,GriStuTerPAN06,StuJPA06,
StuAFdeBr07,GriShiStuTerTro_12LomCon,StuNeutrino_06} we have developed a rather
powerful method for investigation of different phenomena that can appear when
neutrinos and electrons move in background matter. The method discussed is
based on the use of the modified Dirac equations for particles wave functions,
in which the correspondent effective potentials accounting for the standard
model interaction of particles with matter are included. It is similar to the
Furry representation \cite{FurPR51} in quantum electrodynamics, briefly
discussed above. In \cite{StuTerPLB05, GriStuTerPLB05, GriStuTerPAN06,
StuJPA06} we apply the discussed method for elaboration of the quantum theory
of the ``spin light of neutrino" \ ($SL\nu$) in matter. The spin light of
neutrino in matter, one of the four new phenomena  studied in our recent papers
(see for a review \cite{StuNPB05}),  is an electromagnetic radiation that can
be emitted by a massive neutrino (due to its non-zero magnetic moment) when the
particle moves in the background matter. Within the quasi-classical treatment
the existence of this radiation was first proposed and studied in
\cite{Slnu_LobStuPLB03_04_DvoGriStuIJMP05}, while the quantum theory of this
phenomenon was developed in \cite{StuTerPLB05, GriStuTerPLB05, GriStuTerPAN06,
StuJPA06, LobPLB05}.

It should be mentioned here that different forms of the neutrino
quantum wave equations in the presence of matter were used
previously for consideration of modifications of a neutrino
dispersion relation \cite{ManPRD88,NotRaf88,NiePRD89}. As it was
shown in \cite{ManPRD88,NotRaf88,NiePRD89}, the standard result
for the MSW effect \cite{WolPRD78MikSmiYF85} can be derived using
the modified Dirac equation for the neutrino wave function with
the matter potential proportional to the density being added. It
was also shown \cite{ChaZiaPRD88,PanPLB91-PRD92WeiKiePRD97} that a
neutrino energy minimum is at nonzero momentum. Different
interactions of Majorana neutrinos to hypothetical scalar
particles (majorons) in the presence of background matter on the
basis of the modified Dirac equation were studied in
\cite{BerVysBerSmiPLB87_89GiuKimLeeLamPRD92GiuKimLeeLamPRD92BerRosPLB94}.
 The problem
of a neutrino mass generation in different media
\cite{OraSemSmoPLB89,HaxZhaPRD91}, as well as spontaneous
neutrino-pair creation in matter
 were also studied \cite{LoePRL90,KacPLB98,KusPosPLB02,Koe04092549}.

As it has been discussed
\cite{StuJPA06,StuAFdeBr07,GriShiStuTerTro_12LomCon,StuNeutrino_06}, the
approach, developed at first for description of a neutrino motion in the
background matter, can be spread for the case of an electron propagating in
matter. In this Letter we continue the study of an electron motion in matter on
the basis of the modified Dirac equation and its exact solutions. We consider
interaction of an electron with the nuclear matter \cite{Bethe_71} within the
standard model, a problem which has astrophysical relevance (see, for instance,
\cite{KusPosPLB02, Koe04092549}).

As an example of how the developed method works in studies of different
possible processes, generated by electrons propagating in matter, we considered
below in detail electromagnetic radiation that can be emitted by an electron in
the background matter. We have termed this radiation the ``spin light of
electron" \ ($SLe$)  in matter
\cite{StuJPA06,StuAFdeBr07,GriShiStuTerTro_12LomCon,StuNeutrino_06}. It should
be noted here that the term ``spin light of electron"\ was first introduced in
\cite{ITernSPU95} for designation of the synchrotron radiation power particular
contribution connected with an intrinsic magnetic moment of an electron.

Note that our focus is on the standard model interactions of
electrons with the background matter. A similar approach, which
implies the use of the exact solutions of the correspondent
modified Dirac equations, can be developed in the case when
electrons interact with different external fields predicted within
various extensions of the standard model (see, for instance,
\cite{ColKosPRD97_98, ZhuLobMurPRD06}).

\section{Electron quantum states in nuclear matter}

\par

Consider an electron moving in nuclear matter \cite{Bethe_71}. This model of
matter can be used in studies of different processes in astrophysics (see, for
instance, \cite{KusPosPLB02, Koe04092549}). The modified Dirac equation for an
electron with account for matter motion and polarization can be obtained by the
variation procedure applied to the standard Dirac Lagrangian with an additional
effective interaction part
\cite{StuJPA06,StuAFdeBr07,GriShiStuTerTro_12LomCon,StuNeutrino_06})
\begin{equation}\label{Lag_f_e}
\Delta L^{(e)}_{eff}=\tilde{f}^\mu \Big(\bar e \gamma_\mu
{1-4\sin^{2}\theta_{W}+\gamma^5 \over 2} e \Big).
\end{equation}
This leads to the modified Dirac equation,
\begin{equation}\label{ModDirEq_e}
\left\{ i\gamma_{\mu}\partial^{\mu} -\frac{1}{2} \gamma_{\mu}\left( c+\gamma_5\right)
{\tilde{f}}^{\mu} - m_e \right\}\Psi(x)=0,
\end{equation}
where $m_e$ is the electron mass, $c=1-4 \sin^2 \theta_{W}$, and
$\theta_{W}$ is the Weinberg angle. The four-vector
$\tilde{f}^{\mu}$ accounts for the effects of matter motion and
polarization and can be written as
\begin{equation} {\tilde f}^{\mu}=\frac{G_F}{\sqrt
2}(j^{\mu}_n-\lambda^{\mu}_n),
\end{equation}
where  $j^{\mu}_n$  and $\lambda^{\mu}_n$ are the neutron current
and polarization, respectively (for further details see, for
instance, \cite{StuJPA06}).

It should be mentioned here that the form of the obtained modified Dirac
equation (\ref{ModDirEq_e}) for an electron having the standard model
interaction with the neutron matter is similar to the modified Dirac equation
generated in the framework of standard model extensions with CPT violation and
Lorenz breaking \cite{ColKosPRD97_98}. Obviously, the nature of the effective
potential $\tilde{f}^{\mu}$ in our case  is completely different.

In several particular cases the modified Dirac equation
(\ref{ModDirEq_e}) can be solved exactly. We consider below the
case of unpolarized neutrons for which the vector $\tilde f^{\mu}$
is
\begin{equation}\label{f2_e}
\tilde{f}^{\mu}=\frac{G_{F}}{\sqrt{2}}(n_n,n_n{\bf v}),
\end{equation}
where $n_n$ is the number density of the neutron matter and $\mathbf v$ is the
speed of the reference frame in which the mean momentum of neutrons is zero.
Here below we obtain an exact expression for the electron wave function
$\Psi({\bf r},t)$ in a way similar to the one applied  previously in
\cite{GriStuTerPLB05} for solving the problem of a neutrino motion in the
presence of matter background. From Eq.(\ref{ModDirEq_e}) it follows that the
operators of electron momentum, $\hat {\bf p}$, and longitudinal polarization,
${\hat{\bf \Sigma}} {\bf p}/p$, are the integrals of motion, so that, in
particular,
\begin{equation}\label{helicity}
  \frac{{\hat{\bf \Sigma}}{\bf p}}{p}
  \Psi({\bf r},t)=s\Psi({\bf r},t),
 \ \ {\hat {\bm \Sigma}}=
\begin{pmatrix}{\hat {\bm \sigma}}&{0} \\
{0}&{\hat {\bm \sigma}}
\end{pmatrix},
\end{equation}
where $\hat {\bm \sigma}$ are the Pauli matrixes and the values
$s=\pm 1$ specify the two electron helicity states. Assuming the
plane-wave dependence of the electron wave function in matter,
\begin{equation}\label{stat_states}
\Psi({\bf r},t)=e^{-i(  E_{\varepsilon}t-{\bf p}{\bf
r})}u_{\varepsilon, {\bf p},s}({\bf p},E_{\varepsilon}),
\end{equation}
and applying the condition that the equation (\ref{ModDirEq_e})
has a non-trivial solution, we  get the energy spectrum of an
electron moving in the background matter:
\begin{equation}\label{Energy}
  E_{\varepsilon}=\varepsilon{\sqrt{{\bf p}^{2}
  \Big(1-s\alpha_n \frac{m_e}{p}\Big)^{2}
  +m_e^2} +c\alpha_n m_e} ,
\end{equation}
where the matter density parameter $\alpha_n$ is
\begin{equation}\label{alpha}
  \alpha_n=\frac{1}{2\sqrt{2}}{ G}_{F}\frac{n_n}{m_e}.
\end{equation}
The quantity $\varepsilon=\pm 1$ splits the solutions into the two branches
that in the limit of the vanishing matter density, $\alpha_n\rightarrow 0$,
reproduce the positive and negative-energy solutions, respectively, which is
attributed to the particle and anti-particle states. The solution can be
written as
\begin{align}\label{Spinor_AB_e}
\Psi({\bf r},t)= {\displaystyle \frac{e^{-i(
E^{(e)}_{\varepsilon}t-{\bf p}{\bf r})}}{2L^{3/2}}}
\left(%
\begin{array}{c}{\sqrt{1+\frac{m_e}{ {E}_{\varepsilon}-c\alpha_n m_e}}} \ \sqrt{1+s\frac{p_{3}}{p}}
\\
{s \sqrt{1+ \frac{m_e}{ {E}_{\varepsilon}-c\alpha_n m_e}}} \ \sqrt{1-s\frac{p_{3}}{p}}\ \
e^{i\delta}
\\
{  s\varepsilon\sqrt{1- \frac{m_e}{ {E}_{\varepsilon}-c\alpha_n m_e}}} \
\sqrt{1+s\frac{p_{3}}{p}}
\\
{\varepsilon\sqrt{1- \frac{m_e}{ {E}_{\varepsilon}-c\alpha_n m_e}}} \ \
\sqrt{1-s\frac{p_{3}}{p}}\ e^{i\delta}
\end{array}
\right),
\end{align}
where $L$ is the normalization length and $\delta=\arctan({p_2}/{p_1})$, $p_i$
($i=1,2,3$) are the electron momentum components. Note that the difference in
the obtained electron wave function and energy (given by (\ref{Spinor_AB_e})
and (\ref{Energy}) respectively) and the corresponding electron neutrino wave
function and energy \cite{StuTerPLB05} in the neutron background matter is due
to the neutron number density $n$ enters the electron and neutrino
characteristics with opposite signs and the appearance of an additional factor
$c$ in the last term of the electron energy (\ref{Energy}).

\section{Quantum theory of electron spin light in nuclear matter}

\par

As it follows from the expression (\ref{Energy}) for the electron energy in
matter, for a given momentum $p$ the electron energy of the negative-helicity
state exceeds that of the positive-helicity state, thereby enabling the
radiation transition $e_{({-})}\rightarrow e_{({+})} + \gamma$. This process
originates due to the dependence of the electron dispersion law on the density
of matter and may proceed even in the case when the photon refractive index in
matter equals to $n_{\gamma}=1$. We term this radiation the spin light of
electron in matter because it originates from the electron magnetic moment
while the particle moves in the background matter.

  The amplitude of the process is given by $S$ matrix
element
\begin{equation}\label{amplitude_e}
   S_{f i}=-ie \sqrt{4\pi}\int d^{4} x {\bar \psi}_{f}(x)
  ({\gamma}^{\mu}{e}^{*}_{\mu})\frac{e^{ikx}}{\sqrt{2\omega L^{3}}}
   \psi_{i}(x),
\end{equation}
where $-e$ is the electron charge, $\psi_{i}$ and $\psi_{f}$ are the electron
wave functions in the initial and final states respectively, $k^{\mu}=(\omega,
\mathbf{k})$ and ${e}^{\mu}$ are momentum and polarization vectors of the
emitted photon. The further calculations are similar to those performed for the
spin light of neutrino in matter (see
\cite{StuTerPLB05,GriStuTerPLB05,GriStuTerPAN06}). After performing integration
over space-time in Eq.(\ref{amplitude_e}) we get the law of energy-momentum
conservation for the process,
\begin{equation}\label{E_law}
    E=E'+ \omega, \quad \mathbf{p}=\mathbf{p}'+\mathbf{k},
\end{equation}
where unprimed and primed quantities refer to the energy and momentum of the
initial and final electrons, respectively.

 Using the energy-momentum
conservation and the exact expressions for the initial and final electron
energies, which are given by Eq.(\ref{Energy}), we conclude that the only open
channel of the process is the transition with change of the electron helicity
from $s_i=-1$ to $s_f=1$. The revealed asymmetry with respect to the electron
helicity enables us to predict the existence of the electron spin-polarization
effect.

We obtain from (\ref{Energy}) and (\ref{E_law}) for the $SLe$
photon energy
\begin{equation}\label{omega_e}
   \omega=\frac{2\alpha_n m_e p \ [\tilde{E}- (p+\alpha_n m_e)\cos\theta]}
   {(\tilde{E}-p\cos\theta)^2-(\alpha_n m_e)^2},
\end{equation}
where
\begin{equation}\label{E_tilde_e}
     \tilde{E}={E}-c\alpha_n m_e,
\end{equation}
and $\theta$ is the angle between the directions of the radiation
and the initial electron momentum ${\bf p}$. In the case of
relativistic electron and small values of the matter density
parameter $\alpha_n$, that may be realized  for diverse
astrophysical and cosmological environments, for the $SLe$ photon
energy we get from (\ref{omega_e})
\begin{equation}\label{omega_relativistic}
   \omega=
    \frac{1}{\sqrt{2}}G_{F}n_n
    \frac {\beta_e}{1-\beta_e \cos
    \theta},
\end{equation}
where $\beta_e$ is the electron speed. From Eq.(\ref{omega_relativistic}) it
follows that for relativistic electrons the energy range of the $SLe$ may even
extend up to energies peculiar to the spectrum of gamma-rays (see also
\cite{StuTerPLB05,StuJPA06}).

Using expressions for the amplitude (\ref{amplitude_e}) and for
the photon energy (\ref{omega_e}) we obtain the radiation rate
$\Gamma$ and total power ${\mathrm I}$ respectively,
\begin{equation}\label{Gamma_with_Int}
    \Gamma= \frac{e^2}{2}{\int_0}^{\pi} \frac{\omega
   }{1+\tilde{\beta}_e' \, y}  \, S \sin\theta
\, d\theta,
\end{equation}
\begin{equation}\label{I_with_Int}
   {\mathrm I} = \frac{e^2}{2}{\int_0}^{\pi} \frac{\omega^2
   }{1+\tilde{\beta}_e' \, y} S \sin\theta
\, d\theta,
\end{equation}
where
\begin{equation}\label{S}
    \displaystyle  S= (1-y \cos\theta)\left( 1- \tilde{\beta}_e \tilde{\beta}_e'- \frac{{m_e}^2}{\tilde{E}
  \tilde{E}'} \right).
\end{equation}
Here we also introduced the following quantities describing the
initial and final electron respectively,
\begin{equation}\label{beta_e}
   \tilde{\beta}_e=\frac{p+\alpha_n m_e}{\tilde{E}}, \ \ \
   \tilde{\beta}_e'=\frac{p'-\alpha_n m_e}{\tilde{E}'}.
\end{equation}
The electron energy and momentum in the final state are determined by the relations:
\begin{equation}\label{E_prime}
      {E'}=E-\omega, \quad p'=K_e \, \omega - p, \vspace{0.2cm}
\end{equation}
where
\begin{equation}
    K_e={\displaystyle \frac{\tilde{E}- p\cos\theta}{\alpha_n m_e}}, \ \ \
      {\displaystyle y=\frac{\omega-p \cos\theta}{p'}}.
\end{equation}

Performing the integration over the angle $\theta$ in
(\ref{Gamma_with_Int}) and (\ref{I_with_Int}) we get closed
expressions for the radiation rate
\begin{equation}\label{Gamma_tot}
    \Gamma =\frac{e^2m^3}{4p^2}\frac{(1+2a)\left[
(1+2b)^2\ln(1+2b)-2b(1+3b)\right]}{(1+2b)^2\sqrt{1+a+b}},
\end{equation}
and the total radiation power
\begin{equation}\label{I_tot}
     \mathrm{I} =\frac{e^2m^4}{6p^2}\frac{(1+a)\left[
3(1+2b)^3\ln(1+2b)-2b(3+15b+22b^2)\right]-8b^4}{(1+2b)^3},
\end{equation}
where $a=\alpha_n^2+p^2/m_e^2$, $b=2\alpha_n p/m_e$.

As it follows from the above expressions, the $SLe$ rate and total
power are rather complicated functions of the electron momentum
$p$ and the matter density parameter $\alpha_n$. It follows from
Eqs. (\ref{Gamma_tot}) and (\ref{I_tot}) that in the two limiting
cases, $m_e \ll \alpha_n p$ and $m_e \gg \alpha_n p$, expressions
for the rate and power are analytically tractable and the
corresponding much simplified formulas can be obtained. In the
case $\alpha_n\gg m_e/p$ we have:
\begin{equation}\label{m_e<<ap}
\Gamma \approx
\begin{cases}
    \frac{1}{2}e^2\frac{m_e^2}{p}\left[\ln\frac{4\alpha_n p}{m_e}-
    \frac{3}{2}\right], \vspace{0.2cm} \\
    \frac{1}{2}e^2\alpha_n \frac{m_e^3}{p^2}\left[\ln
    \frac{4\alpha_n p}{m_e}-\frac{3}{2}\right],
  \end{cases}
\hspace{-0.3cm}
\mathrm{I} \approx
\begin{cases}
    \frac{1}{2}e^2m_e^2\left[\ln\frac{4\alpha_n p}{m_e}-
    \frac{11}{6}\right], & \text{for} \ \frac{m_e}{p}\ll\alpha_n\ll\frac{p}{m_e},
\vspace{0.2cm}
\\
    \frac{1}{2}e^2\alpha_n^2\frac{ m_e^4}{p^2}\left[\ln\frac{4\alpha_n p}{m_e}-\frac
    {11}{6}\right], & \text{for} \ \alpha_n^{-1}\ll\frac{p}{m_e}\ll\alpha_n.
  \end{cases}
\end{equation}
In the opposite case of $\alpha_n\ll m_e/p$ we get:
\begin{equation}\label{m_e>>ap}
\Gamma \approx
\begin{cases}
    \frac{32}{3}e^2\alpha_n^3  \frac{p^2}{m_e}, \vspace{0.1cm} \\
    \frac{16}{3}e^2\alpha_n^3  p, \vspace{0.1cm} \\
    \frac{32}{3}e^2\alpha_n^4  p,
  \end{cases}
\mathrm{I} \approx
\begin{cases}
    32e^2\alpha_n^4  \frac{p^4}{m_e^2}, & \text{for} \ \alpha_n \ll \frac{m_e}{p} \ll 1, \vspace{0.1cm} \\
    \frac{32}{3}e^2\alpha_n^4  p^2, & \text{for} \ \alpha_n \ll 1 \ll \frac{m_e}{p}, \vspace{0.1cm} \\
    32e^2\alpha_n^6  p^2, & \text{for} \ 1 \ll \alpha_n \ll \frac{m_e}{p}.
  \end{cases}
\end{equation}
The first lines in each case correspond to the radiation of the relativistic
electron. The lines two and three in Eq.(\ref{m_e>>ap}) are for the
non-relativistic case. The remaining line two of Eq.(\ref{m_e<<ap})  describes
the relativistic or non-relativistic case depending on the value of the matter
density.

With the use of the obtained above values of the $SLe$ rate and total power one
can estimate the average emitted photon energy $\left\langle
\omega\right\rangle = \mathrm{I}/{\Gamma}$ for different matter density. In the
case of $\alpha_n\gg m_e/p$ we get from Eq.(\ref{m_e<<ap})
\begin{equation}\label{omega_average 1}
\left<\omega \right> \simeq
\begin{cases}
    \ p, & \text{for} \ \frac{m_e}{p}\ll\alpha_n\ll\frac{p}{m_e},
\vspace{0.2cm}
\\
    \ \alpha_n m_e, & \text{for} \ \alpha_n^{-1}\ll\frac{p}{m_e}\ll\alpha_n,
  \end{cases}
\end{equation}
where it is supposed that $\ln \frac{4\alpha _n p}{m_e} \gg 1$. Thus, for the
relativistic electrons the emitted photons energy is in the range of gamma-rays
what is similar  to the case of the spin light of neutrino
\cite{GriStuTerPLB05,LobPLB05}. Estimations of the initial electron energy
obtained from Eq.(\ref{Energy}) for the two considered in
Eq.(\ref{omega_average 1}) limiting cases show that the photon carries away
nearly the whole of the initial electron energy. This is reminiscent of the
situation has been found in \cite{ZhuLobMurPRD06} for the standard model
extensions.

In the opposite case of $\alpha_n\ll m_e/p$ we get from Eq.(\ref{m_e>>ap})
\begin{equation}\label{omega_average 2}
   \left<\omega\right> \simeq
\begin{cases}
    \ 3\alpha_n\frac{p^2}{m_e}, & \text{for} \ \alpha_n \ll \frac{m_e}{p} \ll 1, \vspace{0.1cm} \\
    \ 2\alpha_n p,  & \text{for} \ \alpha_n \ll 1 \ll \frac{m_e}{p} , \vspace{0.1cm} \\
    \ 3\alpha_n^2p,& \text{for} \ 1 \ll \alpha_n \ll \frac{m_e}{p}.
  \end{cases}
\end{equation}
Estimations of the initial electron energy for the latter three cases give
\begin{equation}\label{enerry_average 2}
   E\simeq
\begin{cases}
    \ p, & \text{for} \ \alpha_n \ll \frac{m_e}{p} \ll 1, \vspace{0.1cm} \\
    \ m_e,  & \text{for} \ \alpha_n \ll 1 \ll \frac{m_e}{p} , \vspace{0.1cm} \\
    \ \alpha_n m_e,& \text{for} \ 1 \ll \alpha_n \ll
    \frac{m_e}{p},
  \end{cases}
\end{equation}
so that small fractions of the initial electron energy are
emitted.

\section{$SLe$ polarization properties}

One of the important features of the $SLe$ is its polarization
properties. It should be mentioned here that in our previous
studies of the polarization properties of the spin light of
neutrino in matter
\cite{StuTerPLB05,GriStuTerPLB05,GriStuTerPAN06} we have shown
that in the case of dense matter the $SL\nu$ photons are
circular-polarized.

We first consider two different linear polarizations of the $SLe$ that are
determined by two orthogonal vectors
\begin{equation}\label{e_12}
  {\bf e}_1= \frac{[{\bm \varkappa}\times {\bf j}]}
  {\sqrt{1-({\bm \varkappa}{\bf j})^{2}}}, \ \
  {\bf e}_2= \frac{{\bm \varkappa}({\bm \varkappa}{\bf j})-{\bf j}}
  {\sqrt{1-({\bm \varkappa}{\bf j})^{2}}},
\end{equation}
where $\bf j$ is the unit vector in the direction of the initial electron
momentum ${\bf p}$. Decomposing the amplitude of the process considered into
contributions from each linear photon polarization, we obtain
\begin{align}\label{I_linear}
    \mathrm{I}^{(1),(2)}  = \frac{e^2}{4}{\int_0}^{\pi}
    \frac{\omega^2}{1+{\tilde\beta}_e^{'}y} \bigg( 1-y\cos & \theta \pm\frac{p}{p^{'}}\sin^2\theta
\bigg)
    \\
    \times & \bigg( 1-{\tilde\beta}_e{\tilde\beta}_e^{'}-\frac{m_e^2}
    {{\tilde E}{\tilde E}'} \bigg)
    \sin\theta d\theta . \notag
\end{align}
It is interesting to investigate these expressions in different limiting cases.
For one particular case determined by the conditions $\alpha_n \ll 1 \ll
\frac{m_e}{p}$, which corresponds to the low matter density, we have
\begin{equation}\label{I linear limiting polariz}
    \mathrm{I}^{(1),(2)} = \left(1\pm \frac{1}{2}\right)\mathrm{I},
\end{equation}
where $\mathrm{I}=\mathrm{I}^{(1)} + \mathrm{I}^{(2)}$. Therefore,
the radiation powers corresponding to the two linear polarizations
differ by a factor of three.
\begin{figure}[t]
\includegraphics[scale=.35]{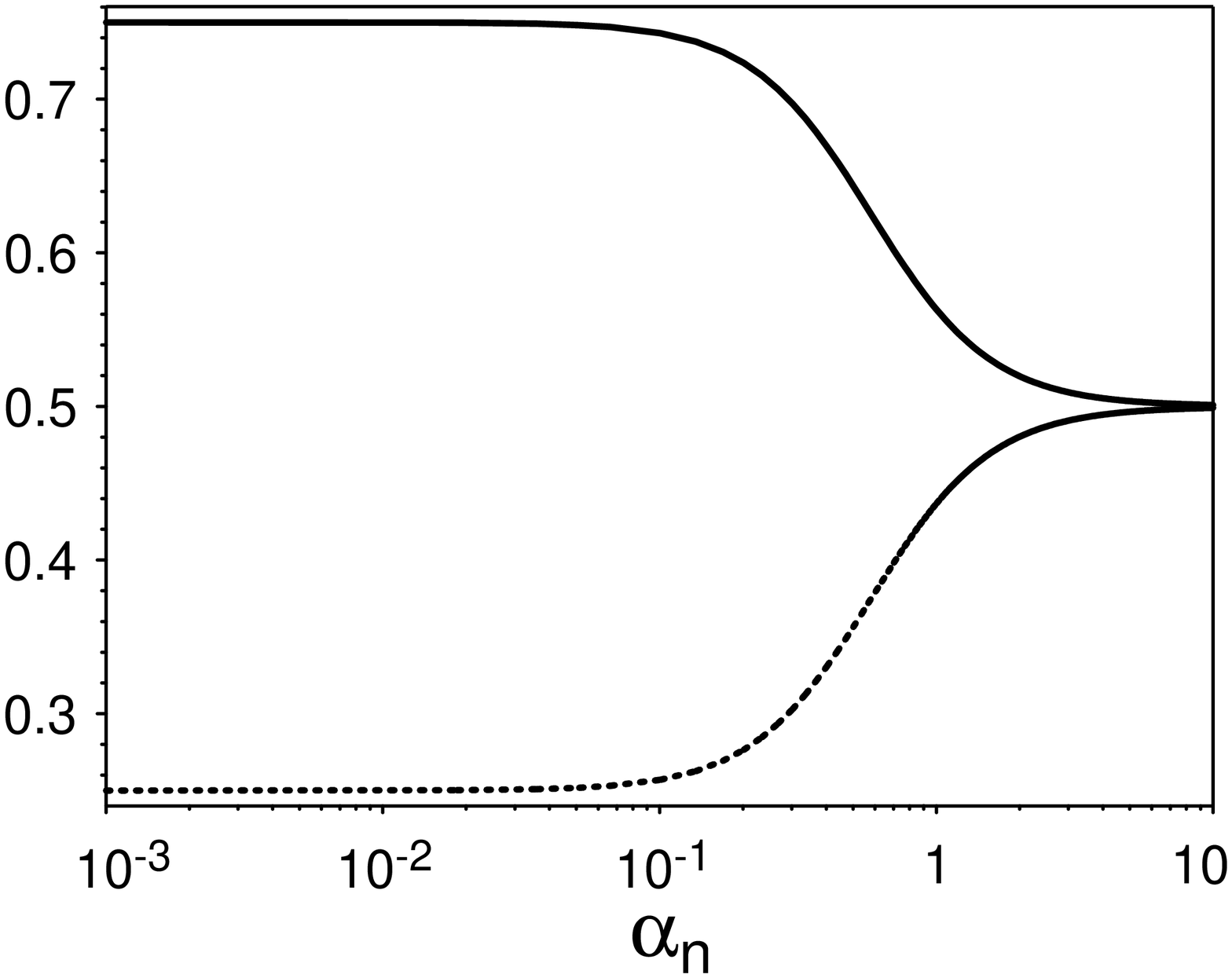}
\hfill
\includegraphics[scale=.35]{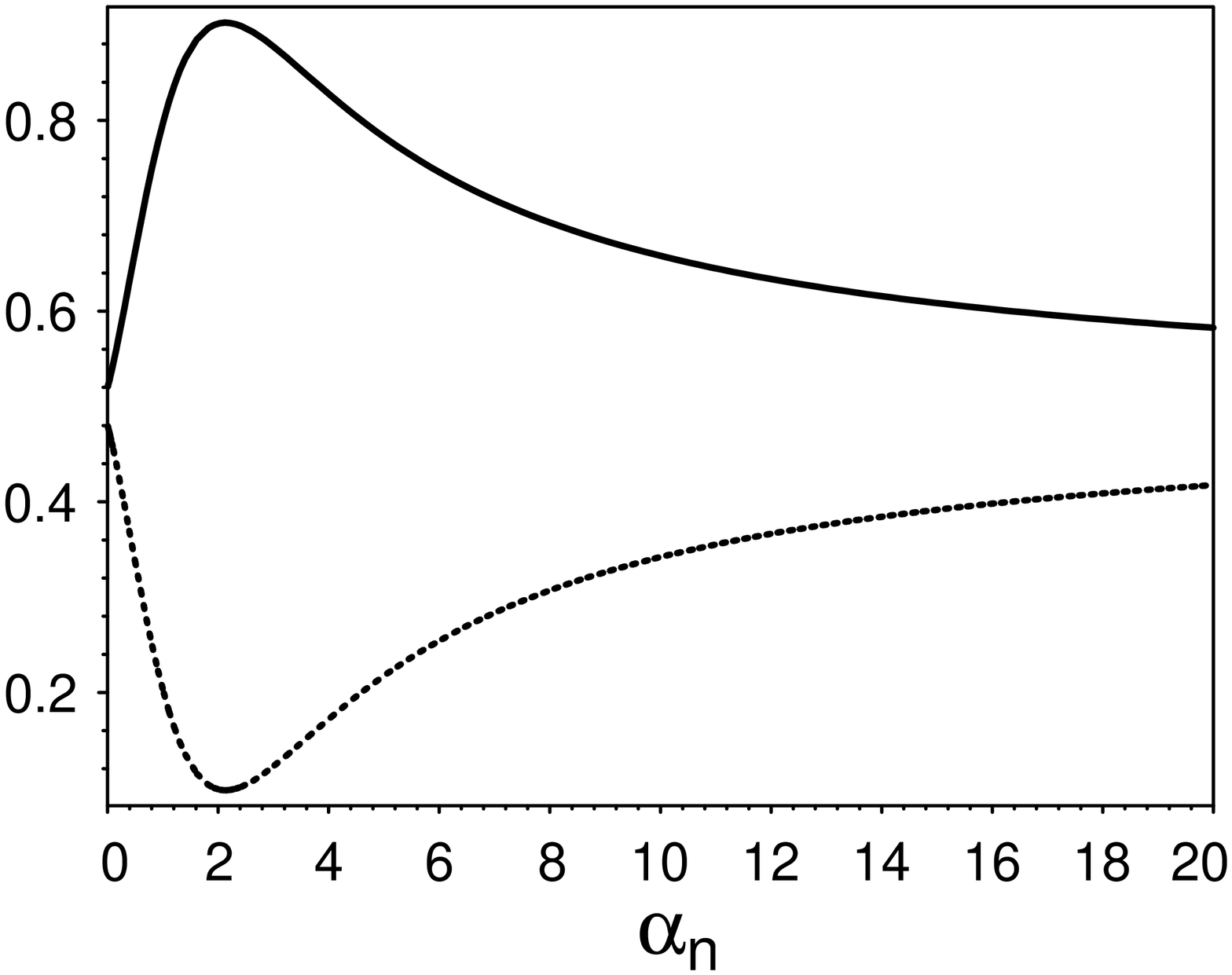}
\\
$\vphantom{a}$ \hspace{3.5cm} $a$ \hspace{8cm} $b$
\\
\vspace{-0.7cm} \caption{Dependence of the $SLe$ linear polarization
contributions $\mathrm{I}^{(1)}$ (solid line) and $\mathrm{I}^{(2)}$ (dashed
line) on the the matter density parameter $\alpha_n$: $(a)$ -- for $p=1 \
\text{keV}$, $(b)$ -- for $p=1 \ \text{MeV}$.}
\end{figure}
In all other cases the radiation powers corresponding to the linear
polarization given by ${\bf e}_1$ and ${\bf e}_2$ are of the same order, so
that the radiation is not polarized:
\begin{equation}\label{I linear limiting no polariz}
    \mathrm{I}^{(1)}\simeq \mathrm{I}^{(2)}\simeq\frac{1}{2}\,
    \mathrm{I}.
\end{equation}
The dependence of the two linear polarization contributions to the
$SLe$ power on the matter density parameter $\alpha_n$ is shown in
Fig.1 for different initial electron momenta $p$. For $p=1 \ keV$
(Fig.1,a) and low values of $\alpha_n$ the degree of linear
polarization is equal to $\mathrm{I}^{(1)}/\mathrm{I}=0.75$. For
$p=1 \ MeV$ (Fig.1,b) the degree of linear polarization gets the
maximal value $\sim 1$ when $\alpha_n\frac{ m_e}{p} \sim 1$.

\begin{figure}[t]
\includegraphics[scale=.35]{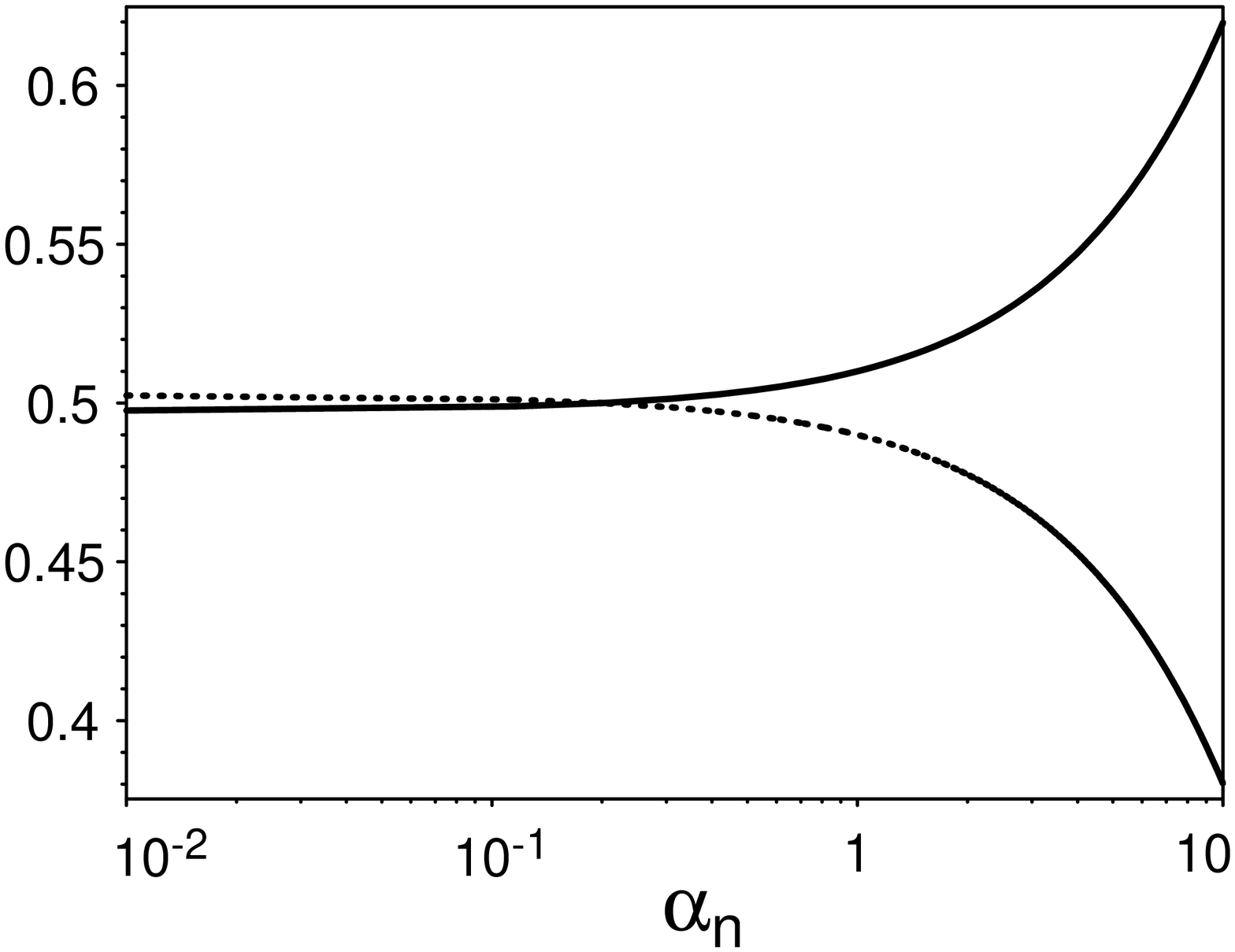}
\hfill
\includegraphics[scale=.35]{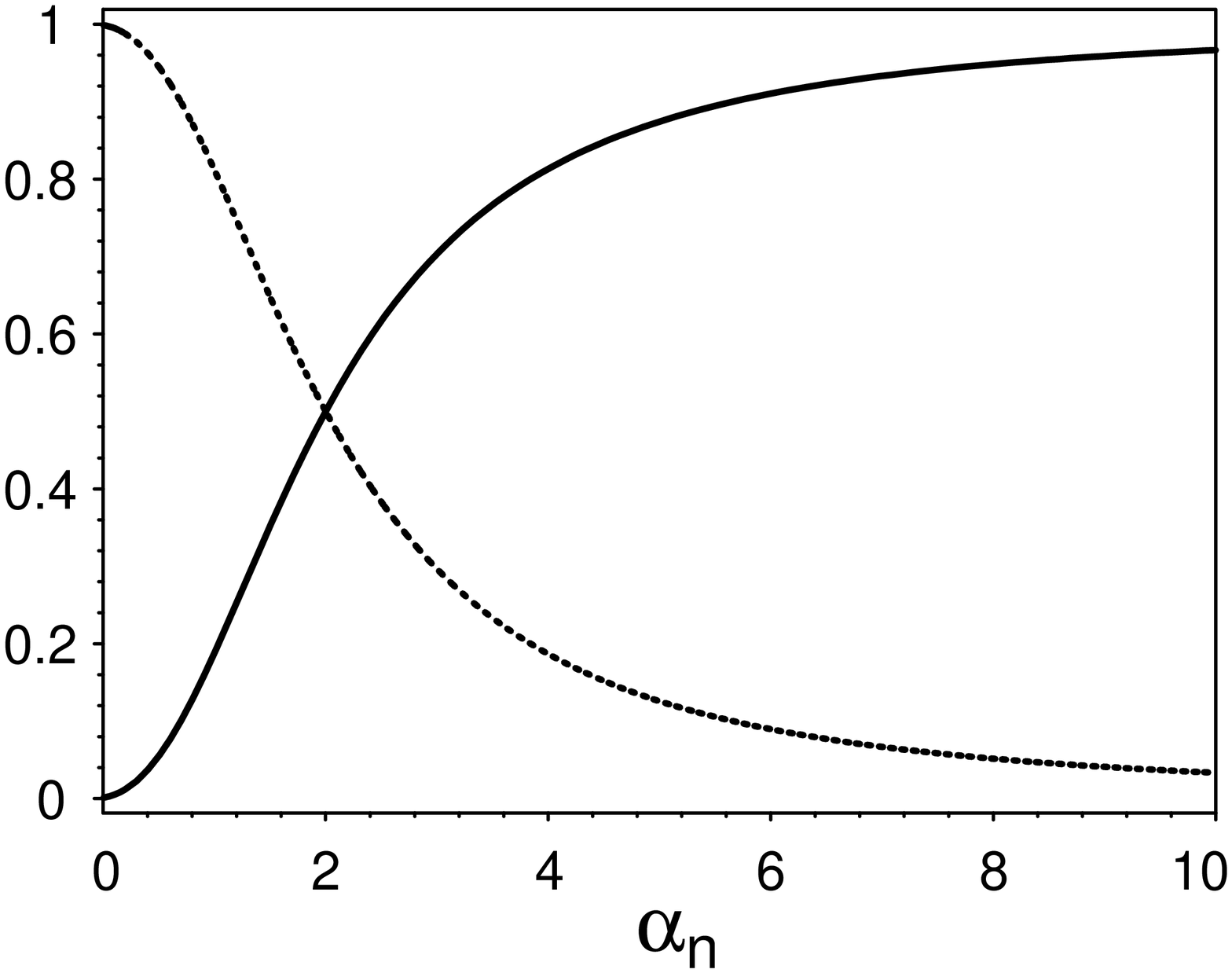}
\\
$\vphantom{a}$ \hspace{3.5cm} $a$ \hspace{8cm} $b$
\\
\vspace{-0.7cm} \caption{Left (solid line) and right (dashed line)
circular polarization contributions as functions of the the matter
density parameter $\alpha_n$: $(a)$ -- for $p=1 \ \text{keV}$,
$(b)$ -- for $p=1 \ \text{MeV}$.}
\end{figure}

Consider the $SLe$ radiation power in the case of the circular
polarization of the emitted photons. As usual, we introduce the
two orthogonal vectors
\begin{equation}\label{circ_pol}
  {\bf e}_{l}=\frac{1}{\sqrt 2}({\bf e}_{1}+il{\bf e}_{2}),
\end{equation}
that are attributed to the two photon circular polarizations with
$l=\pm 1$ for the right and left photon circular polarizations,
respectively.  For the radiation power of the circular-polarized
photons we have
\begin{align}
    \mathrm{I}^{(l)} = \frac{e^2}{4}{\int_0}^{\pi}
    \frac{\omega^2}{1+{\tilde\beta}_e^{'}y} (1+ly)
    ( 1- & l\cos\theta)
    \\
    \times & \bigg( 1-{\tilde\beta}_e{\tilde\beta}_e^{'}-\frac{m_e^2}
    {{\tilde E}{\tilde E}'} \bigg)
    \sin\theta d\theta . \notag
\end{align}
If $\alpha_n\gg m_e/p$, then for the two corresponding subcases we
get that
\begin{equation}
\begin{array}{c}
  \mathrm{I}^{(+1)}\simeq 0, \  \mathrm{I}^{(-1)}\simeq \mathrm{I}, \ \ \ \ \text{for} \ \
\frac{m_e}{p}\ll\alpha_n\ll\frac{p}{m_e},  \vspace{0.1cm}
\\
  \ \mathrm{I}^{(+1)}\simeq \mathrm{I}, \ \mathrm{I}^{(-1)}\simeq 0, \ \ \ \ \text{for} \ \
\alpha_n^{-1}\ll\frac{p}{m_e}\ll\alpha_n.
\end{array}
\end{equation}
In the opposite case $\alpha_n\ll m_e/p$ we have
\begin{equation}
\begin{array}{c}
    \mathrm{I}^{(+1)}\simeq 0, \ \mathrm{I}^{(-1)}\simeq \mathrm{I}, \ \ \ \ \text{for} \ \
\alpha_n \ll \frac{m_e}{p} \ll 1, \vspace{0.1cm}
\\
  \mathrm{I}^{(+1)}\simeq \mathrm{I}^{(-1)}\simeq \frac{1}{2}\,\mathrm{I}, \ \ \ \ \ \, \text{for}
\ \ \alpha_n \ll 1 \ll \frac{m_e}{p}, \vspace{0.1cm}
\\
  \mathrm{I}^{(+1)}\simeq 0, \ \mathrm{I}^{(-1)}\simeq \mathrm{I}, \ \ \ \ \text{for} \ \
1 \ll \alpha_n \ll \frac{m_e}{p}.
\end{array}
\end{equation}

For a wide range of the electron momentum $p$ and the matter
density parameter $\alpha_n$ the radiation is circular polarized,
however the type of polarization (left or right) depends on the
ratio between $\alpha_n$ and $p/m_e$. The dependence of the two
linear polarization contributions to the $SLe$ power on the matter
density parameter $\alpha_n$ is shown in Fig.2 for different
initial electron momenta $p$. For a fixed value of the electron
momentum $p$ the type of polarization changes from the right to
the left with the parameter $\alpha_n$ increase. For $p=1 \ keV$
(Fig.2, a) and small $\alpha_n$ the degree of polarization is very
small, however for big $\alpha_n$ the polarization is maximal. For
$p=1 \ MeV$ (Fig.2, b) and rather small $\alpha_n$ the radiation
is almost right-polarized, for big $\alpha_n$ the type of
polarization changes to the left one. The radiation is unpolarized
when $\alpha_n\frac{ m_e}{p} \sim 1$.

As it can be seen, for rather small values of $\alpha_n$ the $SLe$ is
left-polarized, however degree of polarization decreases with $\alpha_n$
increase and at $\alpha_n \sim \frac{ p}{m_e}$ the $SLe$ is unpolarized. With
the further growth of $\alpha_n$ the right-polarized component dominates and
the degree of polarization increases (with increase of $\alpha_n$).

\section{Conclusion}

We have developed a method for the study of different processes with
participation of electrons subjected to the standard model interactions with
dense matter. This method is based on the use of the modified Dirac equation
for electron wave function in which an effective matter potential is included.
For the nuclear matter composed of neutrons we have found the exact solution of
the modified Dirac equation and determined the electron energy spectrum in
matter.

To illustrate how the developed method works we have elaborated the theory of
electromagnetic radiation of an electron moving in nuclear matter. We have
named this radiation the ``spin light of electron in matter". It is shown, that
for a reasonable values of the matter density the energy range of the $SLe$
photons may even extend up to energies peculiar to the spectrum of gamma-rays.
It has been also shown that the $SLe$ photons can carry away a reasonable
fraction of the initial electron energy and that electron spin polarization
effect can take place. The performed detailed study of the $SLe$ polarization
properties (linear and circular polarizations have been considered) has shown
that for different values of the matter density the radiation can be
significantly polarized and that the type and degree of polarization vary with
change of the electron momentum and density of matter.

Finally, we compare the rates of the spin light of electron and spin light of
neutrino in matter, $\Gamma_{SLe}$ and $\Gamma_{SL\nu}$. In a dense matter with
$n \sim 10^{37}\div 10^{40} \ \text{cm}^{-3}$, for the particles momenta \ $p
\sim 1\div 10^3 \ \text{MeV}$ and for the neutrino mass $m_{\nu}=1 \ \text{eV}$
and magnetic momentum $\mu= 10^{-10}\mu_0$, we have
\begin{equation}\label{ratio_gamma}
R_{\Gamma}=\frac {\Gamma_{SLe}}{\Gamma_{SL\nu}}\sim 10^{16} \div 10^{19},
\end{equation}
in agreement with our previous naive estimation \cite{StuJPA06}. The
corresponding ratio of total power magnitudes for the ${SLe}$ and ${SL\nu}$ is
\begin{equation}\label{ratio_power}
R_{\mathrm{I}}=\frac {\mathrm{I}_{SLe}}{\mathrm{I}_{SL\nu}}\sim 10^{15} \div
10^{19}.
\end{equation}
Considering an electron with momentum $p= 1 \ \text{MeV}$ moving in matter
characterized by the number density $n_n \sim 10^{37} \ \text{cm}^{-3}$ we get
for the rate of the process $\Gamma_{SLe} \sim 3.2 \times 10^{-10} \
\text{MeV}$ which corresponds to the characteristic electron life-time
$T_{SLe}\sim 2\times 10^{-2} \ \text{s}$. Thus, we expect that  the $SLe$ in
matter can be more effective than the $SL\nu$.

The authors are very thankful to Venyamin Berezinsky, Olga Ryazhskaya and Yury
Popov for useful discussions.


\end{document}